\documentclass{pasj00}
\SetRunningHead{Hiroi et al.}{Spectroscopic Observations of GW Librae in
  the 2007 Superoutburst}
\Received{2009/02/04}
\Accepted{2009/04/10}
\Published{}
\begin{document}
\title{Spectroscopic Observations of the WZ Sge-Type Dwarf Nova GW Librae
  during the 2007 Superoutburst}

\author{Kazuo \textsc{Hiroi}$^1$, Daisaku \textsc{Nogami}$^2$,Yoshihiro
  \textsc{Ueda}$^1$, Yuuki \textsc{Moritani}$^1$, Yuichi
  \textsc{Soejima}$^1$, \\
  Akira \textsc{Imada}$^3$, Osamu \textsc{Hashimoto}$^4$, Kenzo
  \textsc{Kinugasa}$^4$,
  Satoshi \textsc{Honda}$^4$, Shin-ya \textsc{Narusawa}$^5$, \\
  Makoto \textsc{Sakamoto}$^5$, Ryo \textsc{Iizuka}$^5$,
  Kentaro \textsc{Matsuda}$^5$, Hiroyuki \textsc{Naito}$^6$, 
  Takashi \textsc{Iijima}$^7$,  and Mitsugu \textsc{Fujii}$^8$}

\affil{$^1$Dept. of Astronomy, Kyoto University, Sakyo-ku, Kyoto 606-8502}
\affil{$^2$Kwasan Observatory, Kyoto University, Yamashina-ku, Kyoto
  607-8471}
\affil{$^3$Dept. of Physics, Kagoshima University, 1-21-35 Korimoto,
  Kagoshima 890-0065}
\affil{$^4$Gunma Astronomical Observatory, 6860-86 Nakayama, Takayama,
  Agatsuma, Gunma 377-0702}
\affil{$^5$Nishi-Harima Astronomical Observatory, Sayo-cho, Hyogo 679-5313}
\affil{$^6$GCOE Office, Nagoya University, Furo-cho, Chikusa-ku, Nagoya, Aichi 464-8602}
\affil{$^7$Astronomical Observatory of Padova, Asiago Section,
  Osservatorio Astrofisico, 36012 Asiago(Vi), Italy}
\affil{$^8$Fujii-Bisei Observatory, 4500 Kurosaki, Tamashima, Okayama
  713-8126}

\KeyWords{accretion, accretion disks --- stars: dwarf novae --- stars:
  individual (GW Librae) --- stars: novae, cataclysmic variables}
\maketitle

\begin{abstract}
  
  We carried out an international spectroscopic observation campaign of
  the dwarf nova GW Librae (GW Lib) during the 2007 superoutburst. Our observation
  period covered the rising phase of the superoutburst, maximum,
  slowly decaying phase (plateau), and long fading tail after the rapid
  decline from the plateau. The spectral features dramatically changed
  during the observations. In the rising phase, only absorption lines of
  H$\alpha$, H$\beta$, and H$\gamma$ were present. Around the maximum, the
  spectrum showed singly-peaked emission lines of H$\alpha$, He~I~5876, He~I~6678,
  He~II~4686, and C~III/N~III as well as
  absorption lines of Balmer components and He~I. 
  These emission lines significantly weakened in the
  latter part of the plateau phase. In the fading tail, all the Balmer
  lines and He~I~6678 were in emission, as observed in quiescence. We find that the
  center of the H$\alpha$ emission component was mostly stable over the
  whole orbital phase, being consistent with the low inclination of the system. 
  Comparing with the observational results of WZ Sge during the
  2001 superoutburst, the same type of stars as GW Lib seen with a high
  inclination angle, 
  we interpret that the change of the H$\alpha$ profile before the
  fading tail phase is attributed to a photoionized region formed
  at the outer edge of the accretion disk, irradiated from the white dwarf 
  and inner disk.
  
\end{abstract}

\section{INTRODUCTION}

Dwarf novae (DNe), a subclass of cataclysmic variables (CVs), are
close binary systems, consisting of a primary star of white dwarf and
a companion star of late-type main-sequence star (for a review, see
\cite{Warner1995,Hellier2001a}).  Mass transferred from the surface of
the companion to the primary by Roche-lobe over flow forms an
accretion disk surrounding the white dwarf. DNe sometimes increase
their luminosities dramatically. This sudden event is called an
outburst, and is currently understood to be caused by the
thermal instability of the accretion disk (for a review, see
\cite{Osaki1996}), although the details of its process are not fully
clarified. In DNe, the accretion disk is the dominant source of
radiation in the optical band during outbursts. This enables us to
investigate basic physics of an accretion disk by optical observations
during outbursts.

SU UMa-type stars are one of the subgroups of DNe, having orbital
periods shorter than 3 hr. These DNe characteristically show two kinds
of outbursts: normal outbursts and superoutbursts. Compared with a
normal outburst, a superoutburst is less frequent, and somewhat
brighter.  In addition, periodic modulations in brightness with a
small amplitude of typically 0.1--0.3 mag, so-called superhumps, are
observed only during superoutbursts.

\begin{figure*}
  \begin{center}
    \FigureFile(160mm,100mm){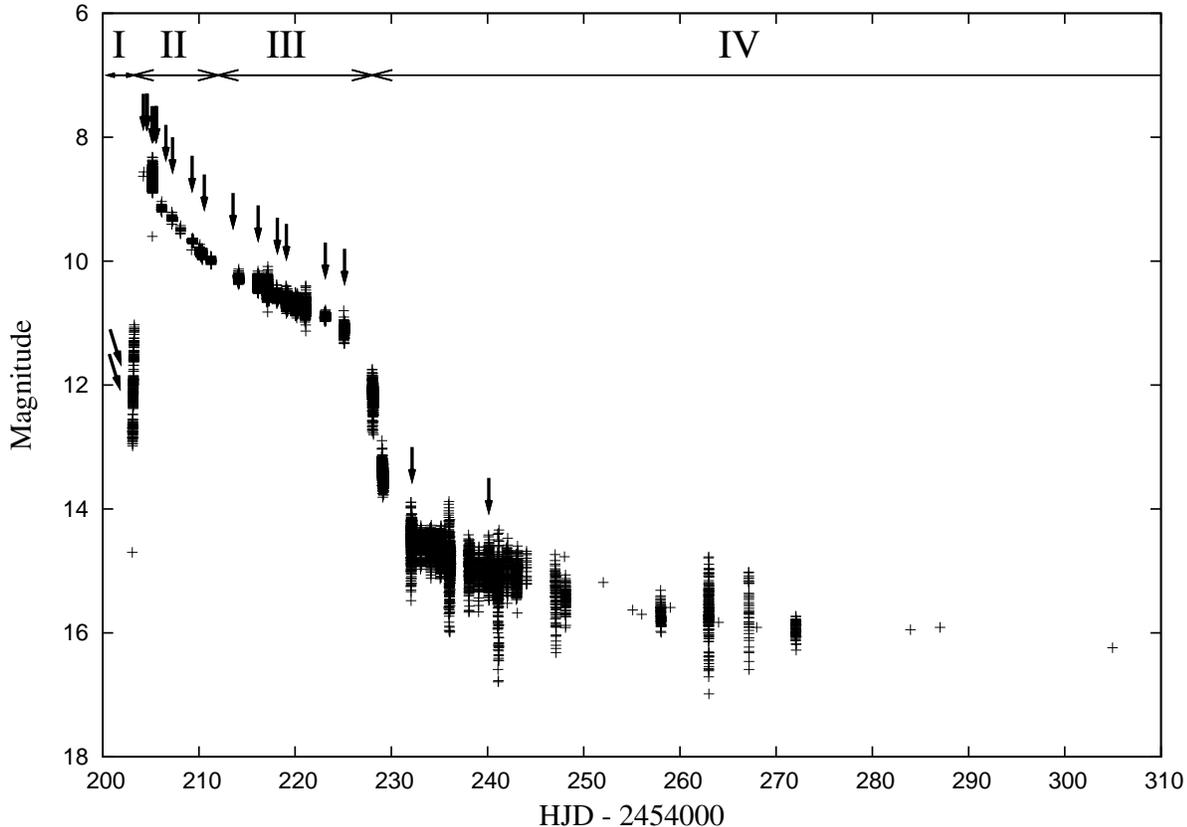}
  \end{center}
  \caption{

    Whole light curve of the 2007 superoutburst in GW Lib generated
    from the data reported to VSNET.  The down-arrows indicate the time
    when the spectra were obtained.  As shown in the top of the figure, we
    separate our observations into four periods to grab the spectral
    features during each period: Period I (April 12), Period II (April 13
    to 19), Period III (April 22 to May 4), and Period IV (May 11 and 19).
    The magnitude of quiescence is about $V=16.8$.
    
  }
  \label{fig:lightcurve}
\end{figure*}

\begin{table*}
  \begin{center}
    \caption{Log of the observations.}
    \label{table:observationlog}
    \begin{tabular}{ccccccc}
      \hline
      Date & UT (Start-End) & Number of        & Exposure & Site
      & Spectral    & HJD (Start-End) \\[-2pt]
      &                & spectra          & time (s) &
      & range (\AA) & (2454000+) \\
      \hline
      April 12 & 15:45-16:45    & 1                & 3600$^*$ &
      Nishi-Harima & 4096-6759   & 203.16113-203.19680 \\
      & 17:42-18:42    & 1                & 3600$^*$ &
      Nishi-Harima & 4096-6759   & 203.24238-203.28405 \\
\phantom{April} 13 & 17:31$^\dagger$      & 1 (3 frame ave.) & 600      &
Fujii-Bisei  & 3752-8329   & 204.23479$^\dagger$ \\
& 25:42-25:47    & 1                & 300      &
Asiago       & 3284-5682   & 204.57578-204.57925 \\
& 25:49-25:59    & 1                & 600      &
Asiago       & 3285-5682   & 204.58064-204.58759 \\
& 26:32-26:37    & 1                & 300      &
Asiago       & 5152-7573   & 204.61051-204.61398 \\
                   & 26:40-25:50    & 1                & 600      &
Asiago       & 5152-7573   & 204.61606-204.62301 \\
\phantom{April} 14 & 13:49-15:57    & 22               & 300      &
Nishi-Harima & 4059-6729   & 205.08067-205.16956 \\
& 16:42-18:53    & 123              & 30       &
Gunma        & 4000-8000   & 205.20081-205.29180 \\
& 24:32-24:42    & 1                & 600      &
Asiago       & 3518-5917   & 205.52722-205.53417 \\
                   & 24:45-24:55    & 1                & 600      &
Asiago       & 3518-5917   & 205.53625-205.54319 \\
& 25:33-25:38    & 1                & 300      &
Asiago       & 5200-7621   & 205.56958-205.57306 \\
& 25:42-25:52    & 1                & 600	  & Asiago
& 5200-7621   & 205.57583-205.58278 \\
\phantom{April} 15 & 25:11-26:09    & 5                & 600	  & Asiago
& 4333-6743   & 206.55436-206.59463 \\
\phantom{April} 16 & 18:28$^\dagger$      & 1 (5 frame ave.) & 600	  &
Fujii-Bisei  & 3753-8311   & 207.27453$^\dagger$ \\
\phantom{April} 18 & 18:57$^\dagger$      & 1 (5 frame ave.) & 600 	  &
Fujii-Bisei  & 3752-8262   & 209.29476$^\dagger$ \\
\phantom{April} 19 & 24:44-25:53    & 3                & 1200     &
Asiago       & 4245-6654   & 210.53579-210.58370 \\
\phantom{April} 22 & 24:38-26:09    & 4                & 1200	  & Asiago
& 4383-6793   & 213.53174-213.59493 \\
\phantom{April} 25 & 14:12-17:12    & 110              & 60	  & Gunma
& 4000-8000   & 216.09711-216.22212 \\
\phantom{April} 27 & 14:28-16:34    & 85               & 60	  & Gunma
& 4000-8000   & 218.10829-218.19579 \\
\phantom{April} 28 & 14:24$^\dagger$	    & 1 (5 frame ave.) & 600      &
Fujii-Bisei  & 3733-8358   & 219.10554$^\dagger$ \\
May 02 & 14:24-16:20    & 80	       & 60	  & Gunma        &
4000-8000   & 223.10564-223.18620 \\
\phantom{May} 04 & 14:24-16:24    & 84               & 60	  & Gunma
& 4000-8000   & 225.10568-225.18901 \\
\phantom{May} 11 & 15:19-15:45     & 6	       & 180	  & Gunma        &
4000-8000   & 232.14396-232.16202 \\
\phantom{May} 19 & 14:55-15:31     & 6	       & 300      & Gunma
& 4000-8000   & 240.12731-240.15231 \\
\hline
\multicolumn{7}{@{}l@{}}{\hbox to 0pt{\parbox{180mm}{\footnotesize
      
      \par\noindent
      \footnotemark[$^*$] It was very hazy.
      \par\noindent
      \footnotemark[$^\dagger$] The average of the midst of the exposure
      times of each frame.
    }\hss}}
    \end{tabular}
  \end{center}
\end{table*}

A subset of SU UMa-type stars that has very long recurrence intervals
of the outburst ($>$5 years) is called a WZ Sge-type star, and shows
only superoutbursts (\cite{Kato2001}, and references therein).  This
type of stars has the shortest orbital periods of $\approx$80 min
among DNe, and is supposed to have quite low mass-transfer rates
(e.g., \cite{Osaki1995}).
The currently most debated issues regarding WZ Sge
stars are the origin of early superhumps and rebrightenings. The early
superhumps are observed at the very early phase of superoutbursts in
WZ Sge stars before emergence of ordinary superhumps (see e.g.,
\cite{Kato1996, Ishioka2002, Osaki2002, Patterson2002, Nogami2007}).
The rebrightening(s) often occurs a few to several days after the end
of the plateau phase of the superoutburst in WZ Sge stars (and some of
SU UMa stars with short orbital periods), although the recurrence
cycle of the outburst is quite long, as mentioned above (see e.g.,
\cite{Nogami1997, Osaki2001, Ishioka2001, Patterson2002, Kato2004a}).

GW Lib was discovered during an outburst in 1983, and initially
classified as a nova because of its large outburst amplitude
(\cite{Gonzalez1983}; \cite{Duerbeck1987}).  Subsequent spectroscopic
observations of GW Lib, however, showed that strong and narrow
emission
components superposed on broad Balmer absorption lines were present
(\cite{Duerbeck&Seitter}; \cite{Ringwald1996}; \cite{Szkody2000}).
This leads to the conclusion that GW Lib is a DN with a
remarkably low mass-transfer rate\footnote{From early on, some amateurs 
  suggested that GW Lib was a DN with a large-amplitude outburst 
  (Henkousei 109 (in Japanese)), and were monitoring it.
  This fact helped the rapid discovery of the 2007 superoutburst and the prompt
  observations.}.

In fact, GW Lib has an exceptionally short orbital period of $\sim$ 77 min
(\cite{Thorstensen2002}), which is nearly equal to the observed period
minimum of normal hydrogen-rich CVs (see e.g. \cite{Patterson2003,
  Pretorius2007}).  The inclination of GW Lib was estimated to be
$\sim$11$^\circ$ from the comparison of the emission-line width with
WZ Sge (\cite{Thorstensen2002}). Trigonometric parallax and proper
motions indicate that the distance of GW Lib is 104 pc
(\cite{Thorstensen2003}). One of the most curious properties of GW Lib
is that it is the first cataclysmic variable revealed to show
non-radial photometric pulsations of the primary in optical
(\cite{Warner1998}; \cite{vanZyl2000}; \cite{Woudt2002};
\cite{vanZyl2004}), although these pulsations were not detected in the
long fading tail of the 2007 superoutburst (\cite{Copperwheat2009}). 
The primary star of GW Lib is, hence, thought to
be a ZZ Cet-type white dwarf.  These pulsations are also present in UV
(\cite{Szkody2002}), but not observed in X-rays (\cite{Hilton2007}).

At 2007 April 12.494 (UT), we received a VSNET message by R.Stubbings
(vsnet-alert 9279) reporting that GW Lib was in outburst for the first
time in 24 years since the discovery (for VSNET, see
\cite{Kato2004b}). Following this alert, we conducted an international
spectroscopic observation campaign of GW Lib.  Our observations
started from the rising phase of this outburst (\cite{Narusawa2007}).
This is the second
successful campaign of such intensive spectroscopic observations of a
WZ Sge-type DN, next to the one on WZ Sge during the 2001
superoutburst \citep{Nogami2004}. Photometric observations during this
outburst detected superhumps, confirming that this outburst is a
superoutburst, and that GW Lib is really a member of WZ Sge-type DNe 
(\cite{Kato2008}; Imada et al., in preparation; Uemura et al., in preparation).

Section~\ref{section:OBSERVATION} describes our observational
protocols and log. We present the optical spectra obtained in our
observations in section~\ref{section:RESULT}, and discuss 
the evolution of the 
accretion disk structure during the superoutburst
in section~\ref{section:DISCUSSION}.
A summary is given in section~\ref{section:SUMMARY}.

\section{OBSERVATION}
\label{section:OBSERVATION}

Four observatories participated
in this campaign: Asiago Astrophysical
Observatory, Gunma Astronomical Observatory, Nishi-Harima Astronomical
Observatory, and Fujii-Bisei Observatory. The Asiago Astrophysical
Observatory is located in Italy, and the others are in
Japan. Figure~\ref{fig:lightcurve} shows the whole light curve of GW
Lib during the 2007 superoutburst, generated from the data reported to
the VSNET. As seen in figure~\ref{fig:lightcurve}, our observations
covered the time from the rising phase of the superoutburst to the
long fading tail. All of the obtained spectral data were reduced in a
standard way using the IRAF package. Table~\ref{table:observationlog}
gives a journal of the observations.

\subsection{Asiago Astrophysical Observatory}

On 2007 April 13, 14, 15, 19, and 22, we took the spectra of a medium
resolution (R $\approx$ 1000) with a Boller \& Chivens spectrograph
mounted on the 122-cm Galileo telescope of the Asiago Astrophysical
Observatory, using a 512 $\times$ 512-pixel CCD detector.  The
integration times varied from 300 s to 1200 s, depending on the
conditions.

\subsection{Gunma Astronomical Observatory}

At the Gunma Astronomical Observatory, observations were performed on
2007 April 14, 25, 27, May 2, 4, 11, and 19.  The spectra of a low
resolution (R $\approx$ 400 -- 500) were obtained with the Gunma LOW
resolution Spectrograph (GLOWS) mounted on a 150-cm Ritchey-Chretien
telescope. To examine variability of the source on a short time-scale,
we adopted relatively short integration times (30 s or 60 s),
except for May 11 and 19.

\subsection{Nishi-Harima Astronomical Observatory}

We took the spectra of a low resolution (R = 370 at H$\alpha$) with
the optical spectrograph, MALLS, mounted on the 2.0-m NAYUTA telescope
of the Nishi-Harima Astronomical Observatory on 2007 April 12 and 14.
The integration times were 3600 s and 300 s on April 12 and 14,
respectively.

\subsection{Fujii-Bisei Observatory}

Observations were performed on 2007 April 13, 16, 18, and 28 at the
Fujii-Bisei Observatory.  The low resolution (R = 600 at 5852 \AA)
spectra were taken with the spectrograph, FBSPEC-2, mounted on a 28-cm
Schmidt-Cassegrain telescope with an integration time of 600 s.

\section{RESULT}
\label{section:RESULT}

\begin{figure}
  \begin{center}
    \FigureFile(80mm,50mm){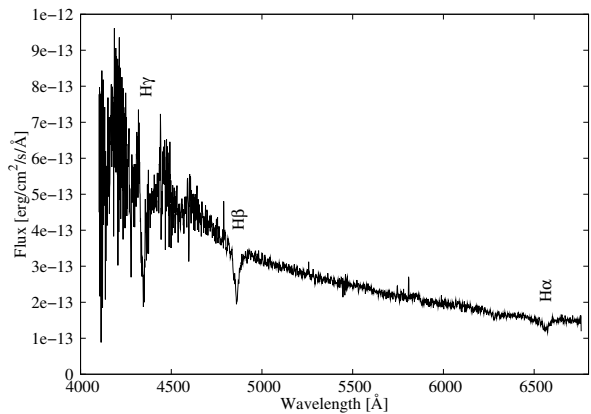}
  \end{center}
  \caption{
    Spectrum obtained at the Nishi-Harima Astronomical Observatory on
    2007 April 12 in the rising phase of the superoutburst.
  }
  \label{fig:Period-I_cal}
\end{figure}

Figure \ref{fig:Period-I_cal} shows the spectrum obtained on 2007
April 12, just in the rising phase of the superoutburst. We can see
the blue continuum, on which deep Balmer absorption lines (H$\alpha$,
H$\beta$, and H$\gamma$) are superposed. There are no other prominent
lines. This profile is characteristic of DNe in outburst.

\begin{figure*}
  \begin{center}
    \FigureFile(160mm,100mm){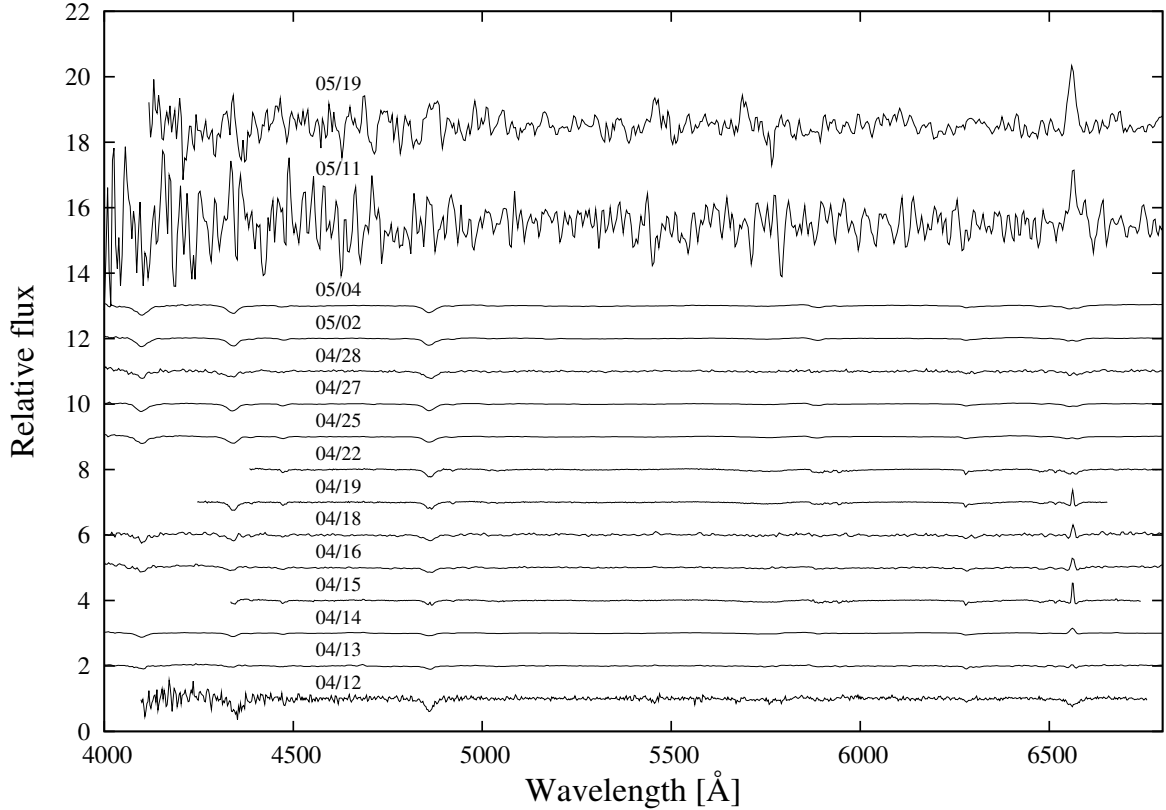}
  \end{center}
  \caption{
    Representative normalized spectra of each day from 2007 April 12 to
    May 19. For visuality, the data after April 13 are vertically shifted;
    the offsets are +1 to +12 (with a step of 1) for April 13 to May 4,
    +14.5 for May 11, and +17.5 for May 19.
  }
  \label{fig:profile-variation}
\end{figure*}

Figure~\ref{fig:profile-variation} displays all the ``normalized''
representative spectra of each day from 2007 April 12 to May 19.  It
is seen that H$\beta$, H$\gamma$, and H$\delta$ were consistently in
absorption during the observations, except for during the fading tail
(May 11 and 19). On the other hand, the profile of H$\alpha$
dramatically changed. It was an absorption line in the rising phase of
the superoutburst. Around the superoutburst maximum, it became an
emission line, while it was again in absorption with a weak emission
component in the latter part of the plateau phase. It finally turned
to a strong emission line in the long fading tail.  In addition to
Balmer lines, many He lines were present in emission or in absorption.
The high-excitation emission lines of He~II~4686 and C~III/N~III were
observed only around the superoutburst maximum.

We separate the whole observations into four periods, and examine the
spectral features in each period. As denoted in
figure~\ref{fig:lightcurve}, Period I is the rising phase (April 12),
Period II is the former part of the plateau phase (April 13 to 19),
Period III is the latter part of the plateau phase (April 22 to May
4), and Period IV is the long fading-tail phase (May 11 and 19). The
maximum of the superoutburst belongs to Period II.  Note that early
superhumps were not observed in this system, and the genuine
superhumps emerged around 2007 April 21, reaching its maximum
amplitude on 2007 April 23 in Period III (see Uemura et al., in
preparation; Imada et al. in preparation).

\subsection{Period I: The rising phase (April 12)}

\begin{figure}
  \begin{center}
    \FigureFile(80mm,50mm){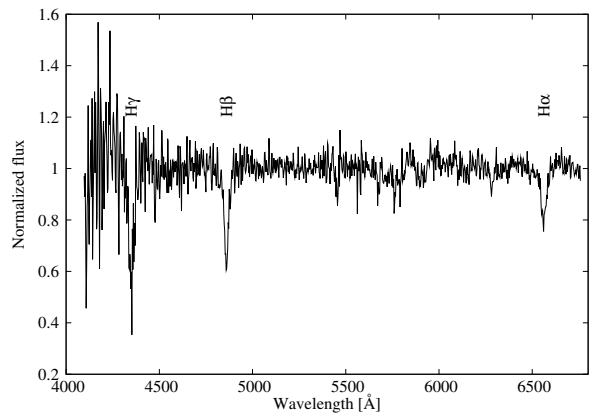}
  \end{center}
  \caption{

    Normalized spectrum taken at the Nishi-Harima Astronomical
Observatory on 2007 April 12 in Period I.

  }
  \label{fig:Period-I}
\end{figure}

Figure~\ref{fig:Period-I} shows the normalized spectrum on 2007
April 12 in Period~I, during the rising phase, derived from the raw
spectrum shown in figure~\ref{fig:Period-I_cal}. Only Balmer
absorption lines of H$\alpha$, H$\beta$, and H$\gamma$ were seen
without any significant wavelength shifts.

\subsection{Period II: The former part of the plateau phase (April 13 to 19)}

\begin{figure}
  \begin{center}
    \FigureFile(80mm,50mm){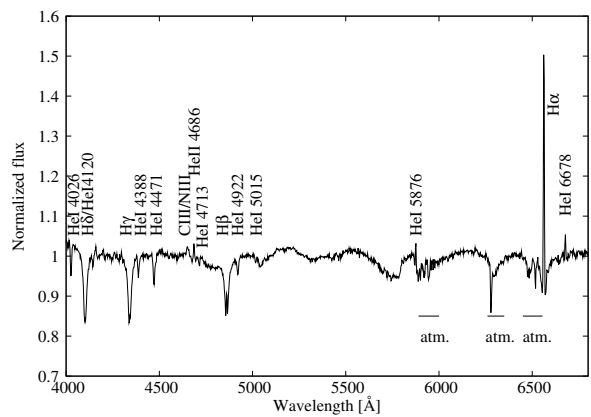}
  \end{center}
  \caption{
    
    Normalized spectrum in Period II taken on 2007 April 14 at the
    Asiago Astrophysical Observatory. It is combined from bluer and
    redder ones.
    
  }
  \label{fig:Period-II}
\end{figure}

Figure~\ref{fig:Period-II} shows the normalized spectrum obtained on
2007 April 14 in Period II. This is a daily-averaged spectrum,
combined of bluer and redder ones. H$\alpha$ turned to a strong
emission line from an absorption one in Period I.  Although other
Balmer lines still remained in absorption, H$\beta$ and H$\gamma$ were
accompanied by a weak emission component.  We can also see that there
were many He~I absorption lines: He~I 4026, 4120, 4388, 4471, 4713,
4922, and 5015.  On the other hand, He~I~5876, He~I~6678 and high-excitation lines
of He~II~4686 and C~III/N~III were in emission.  They, however, gradually
disappeared in the latter of this period. All the emission lines were
singly-peaked, being consistent with the low inclination of this
system.

\subsection{Period III: The latter part of the plateau phase (April 22 to May 4)}

\begin{figure}
  \begin{center}
    \FigureFile(80mm,50mm){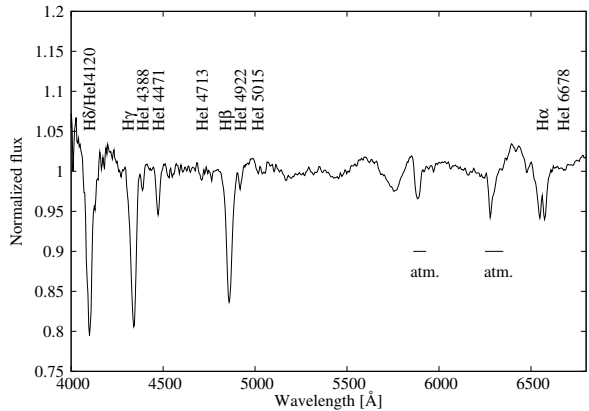}
  \end{center}
  \caption{
    
    Normalized spectrum in Period III, obtained on 2007 April 25 at the
    Gunma Astronomical Observatory.
    
  }
  \label{fig:Period-III}
\end{figure}

The normalized spectrum taken on 2007 April 25, a representative day
in Period III, is shown in figure~\ref{fig:Period-III}. This is the
average of all the spectra obtained at the Gunma Astronomical
Observatory on that day. 
It is seen that H$\alpha$ turned to an absorption line with a narrow emission component,
although it was in emission in Period II.
He~I~6678 also changed from an emission line into a very weak absorption one.
We cannot clearly see the He~I~5876 feature because of the low
wavelength resolutions of the spectra obtained in this period.
Other Balmer lines were still in absorption,
but became deeper than in Period II.  The emission components of H$\beta$
and H$\gamma$ observed in Period II disappeared. The He~I lines,
except for He~I~5876 and He~I~6678, still remained in absorption.

\subsection{Period IV: The long fading tail phase (May 11 and 19)}

\begin{figure}
  \begin{center}
    \FigureFile(80mm,50mm){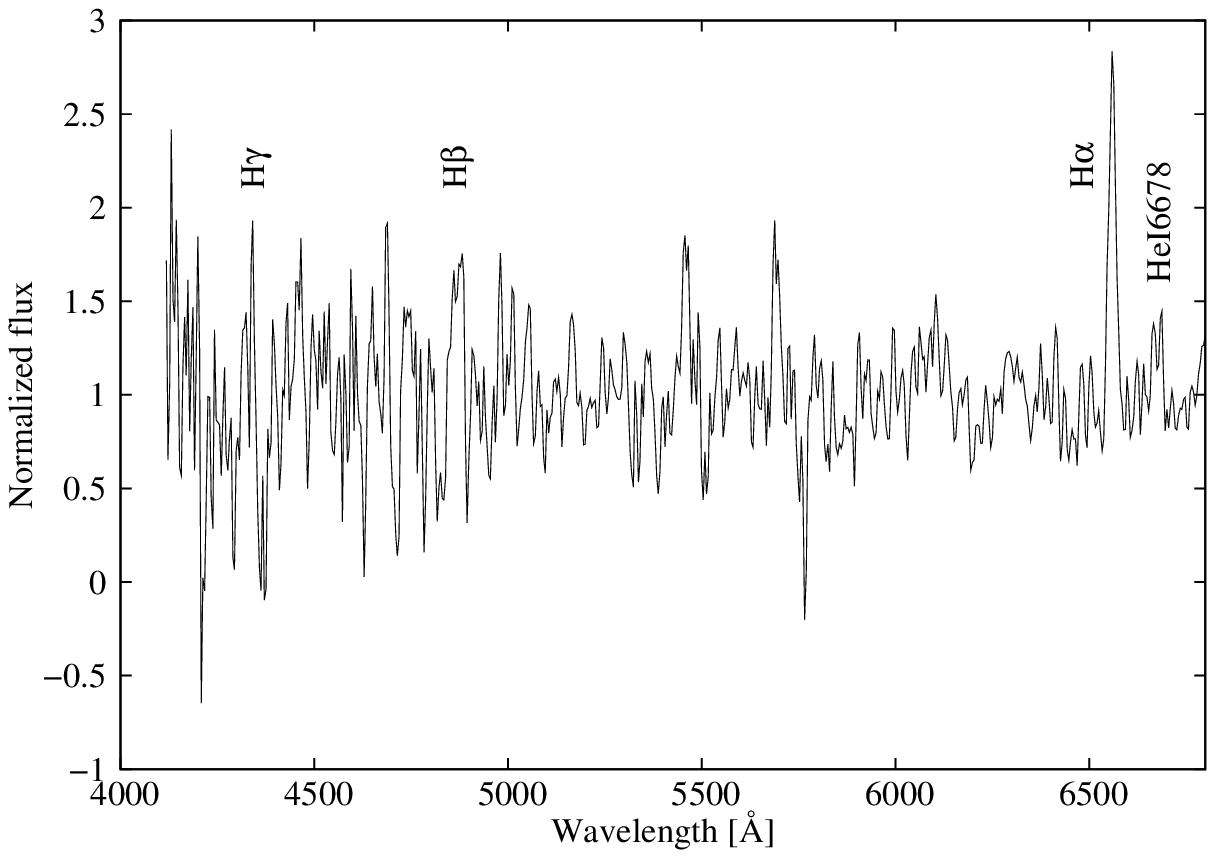}
  \end{center}
  \caption{
    
    Normalized spectrum obtained at the Gunma Astronomical Observatory
    on 2007 May 19 in Period IV.  
    
  }
  \label{fig:Period-IV}
\end{figure}

Figure \ref{fig:Period-IV} shows the normalized spectrum on 2007 May
19 in Period IV. This is the averaged spectrum taken at the Gunma
Astronomical Observatory on that day. In this period, the
signal-to-noise ratio was low due to the faintness of GW Lib.  We can
see, however, that H$\alpha$, H$\beta$, H$\gamma$, and He~I~6678 were
present in emission. Note that the FWHM of H$\alpha$ was $\approx$22
\AA\ on 2007 May 19, which is 1.5 times broader than that on 2007
April 14, $\approx$13 \AA, just after the superoutburst maximum (both
data were obtained with the same instrument at the Gunma Astronomical
Observatory and hence direct comparison can be made.).

\section{DISCUSSION}
\label{section:DISCUSSION}

\subsection{Variation of Emission-Line Center as a Function of Orbital Phase}

\begin{figure*}
  \begin{center}
    \FigureFile(50mm,50mm){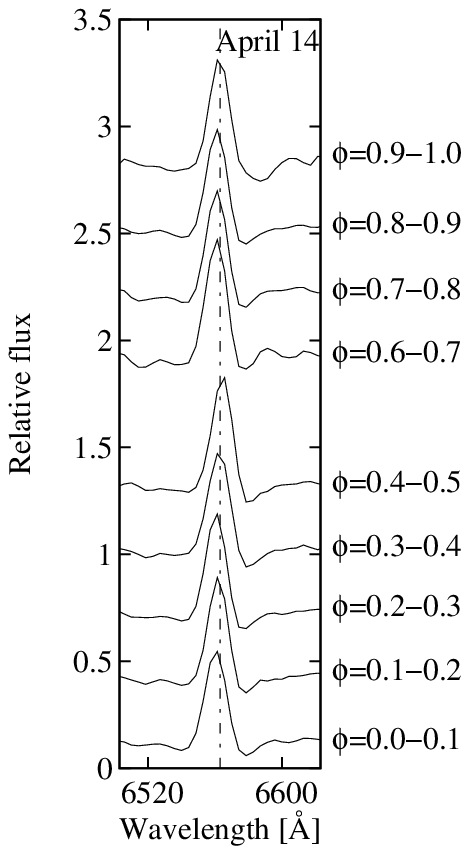}
    \FigureFile(50mm,50mm){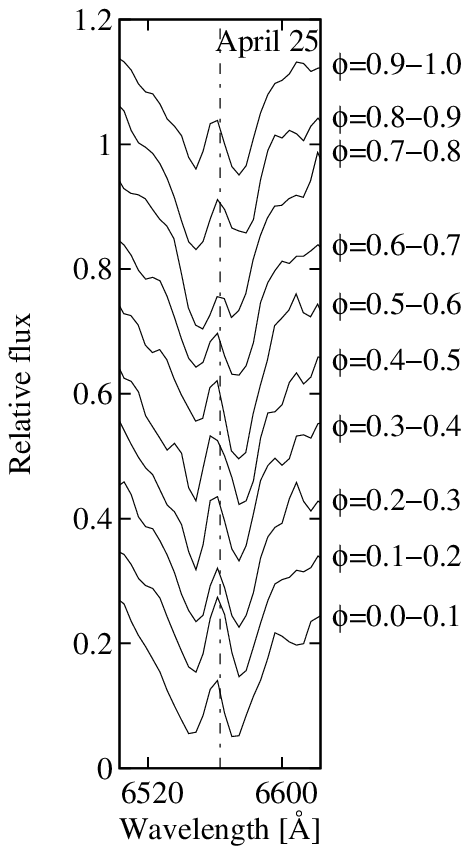}
    \FigureFile(50mm,50mm){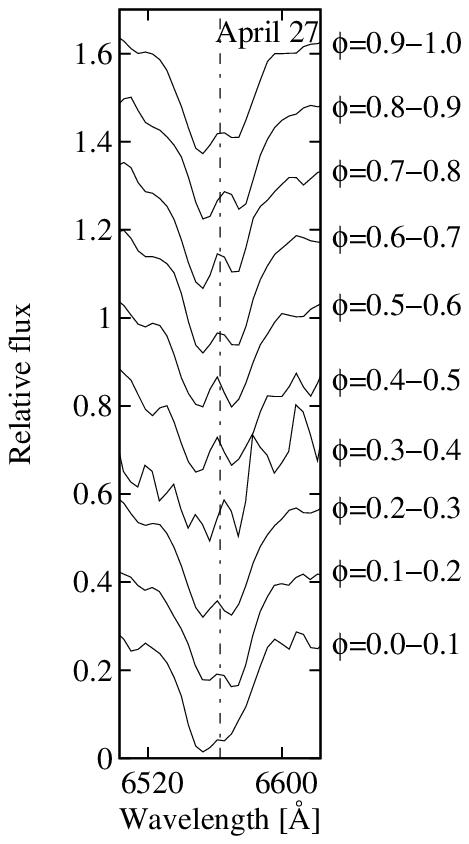}
    \FigureFile(50mm,50mm){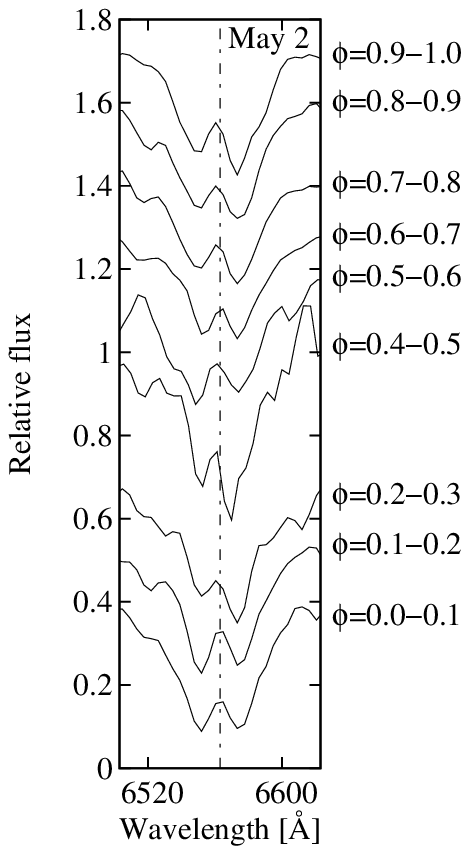}
    \FigureFile(50mm,50mm){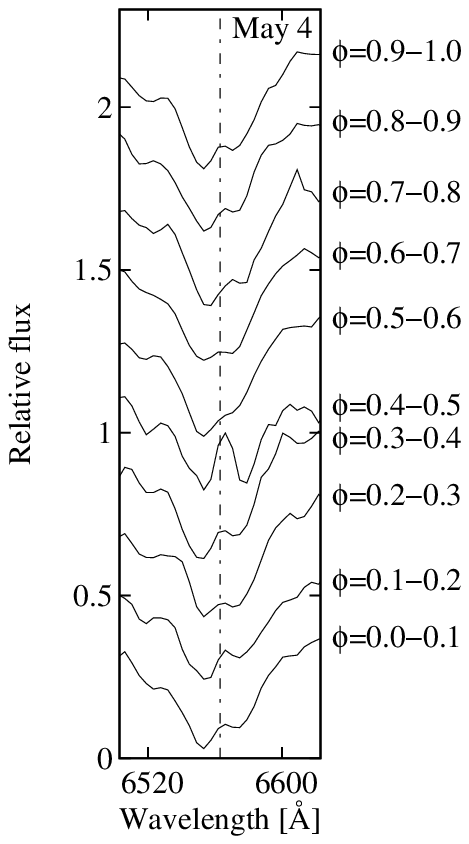}
  \end{center}
  \caption{
    
    H$\alpha$ profile variations as a function of orbital
    phase on April 14, 25, 27, May 2, and 4.  The dot-dashed line
    corresponds to the rest-frame wavelength of H$\alpha$. There is no
    clear variation of the central wavelength of the emission 
    component with orbital phase within our low-resolution spectra.
    
  }
  \label{fig:Ha-phase-variation}
\end{figure*}

Using the H$\alpha$ emission components on 2007 April 14, 25, 27, May 2,
and 4, we examine the profile variation as a function of orbital 
phase of GW Lib.
We calculate the average profiles of H$\alpha$ in bins of
$\Delta$ $\phi$ = 0.1.
The orbital phase, $\phi$, is defined by the following equation:
\begin{equation}
  \phi = \frac{HJD - T_0}{P_{orb}} - E_0,
\end{equation}
where $ T_0 = 2451340.6580 $, $ P_{orb} = 0.05332 $ (\cite{Thorstensen2002}), 
and $E_0$ is an integer that gives $0<\phi<1$ at the observation start time on 
each day.
We do not have any data
in $\phi$ = 0.5--0.6 and 0.3--0.4 on April 14 and May 2, respectively,
and hence the spectra in those phase are not available.

Figure~\ref{fig:Ha-phase-variation} is the result of each day.
The dot-dashed line shows the rest-frame wavelength
of H$\alpha$. As can be seen in the figure, the center of the emission
component is almost stable over the whole orbital phase 
on each day. 
This result is consistent with the low
inclination of the system ($\sim 11^\circ$). The depths of blue and
red troughs besides the emission component do not vary with the
orbital phase, either, within the quality of our spectra.

\subsection{Structure of the Accretion Disk during Superoutburst}

\begin{figure*}
  \begin{center}
    \FigureFile(45mm,80mm){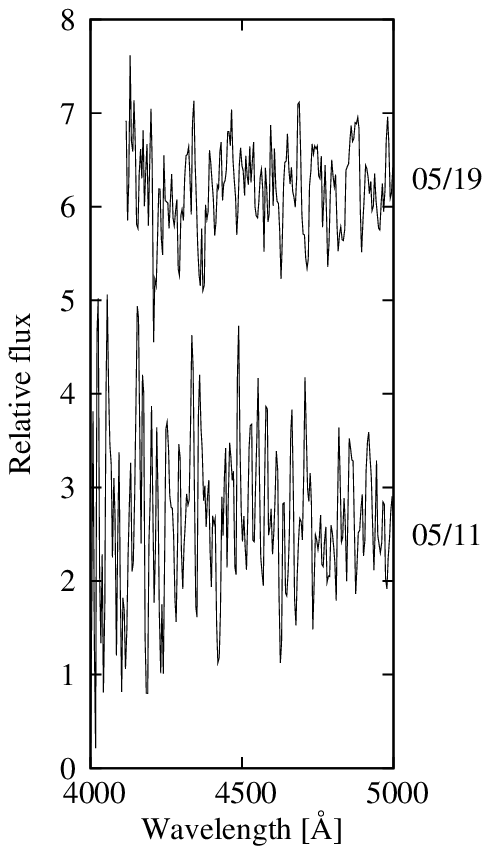}~~~~~~~~~~~~~~~~~~~~
    \FigureFile(45mm,80mm){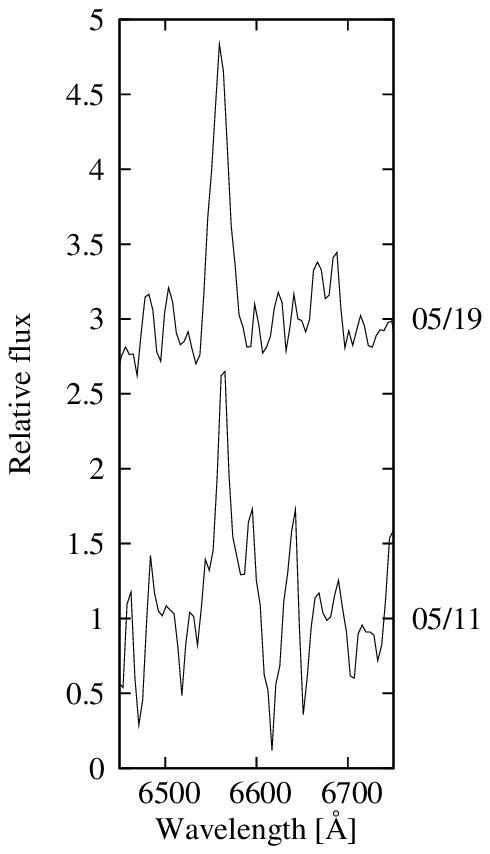}\\
    \FigureFile(45mm,80mm){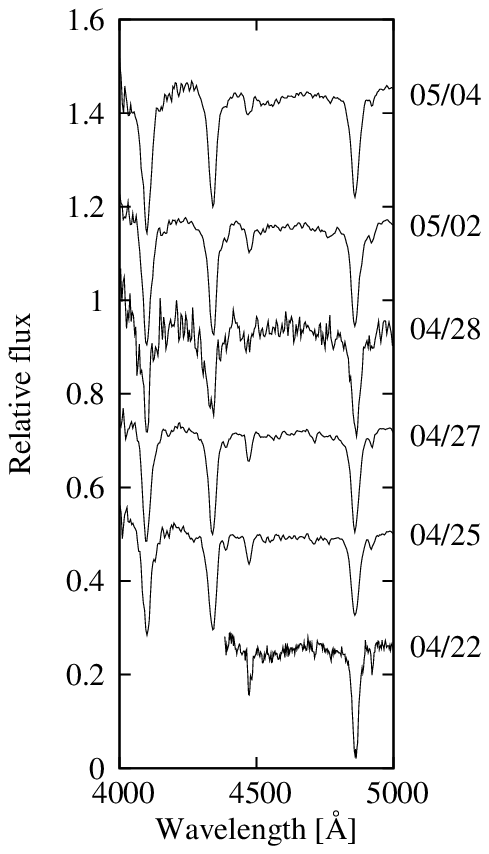}~~~~~~~~~~~~~~~~~~~~
    \FigureFile(45mm,80mm){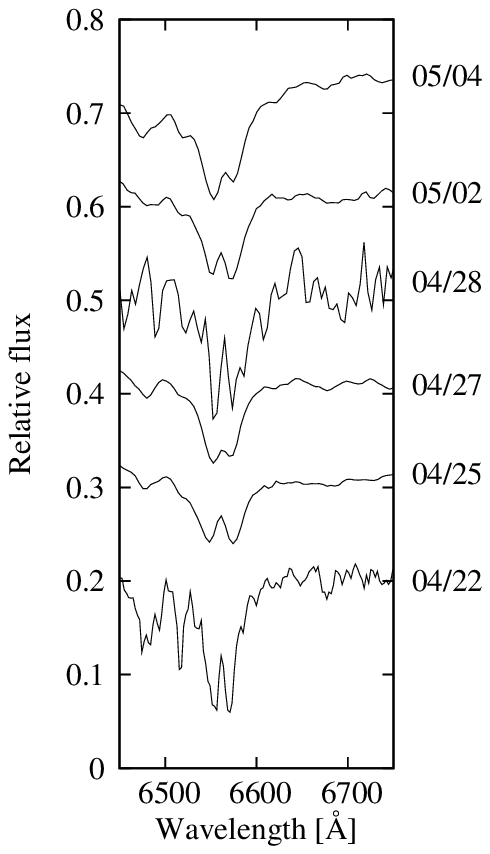}\\
    \FigureFile(45mm,80mm){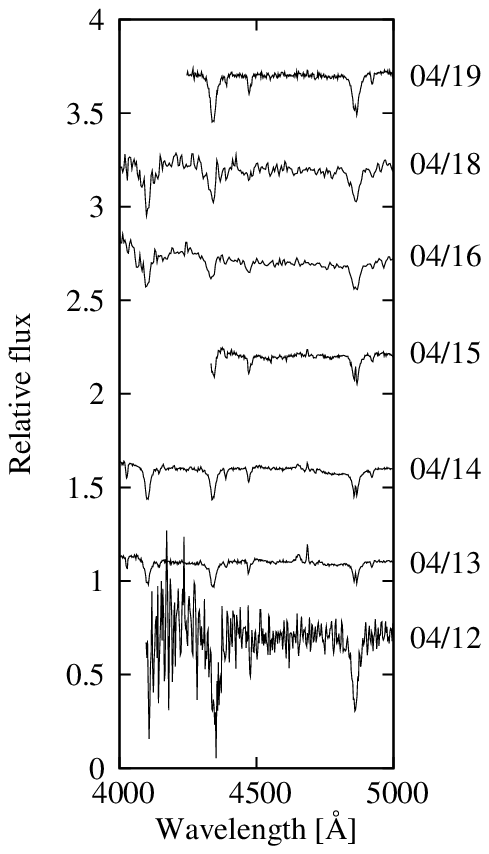}~~~~~~~~~~~~~~~~~~~~
    \FigureFile(45mm,80mm){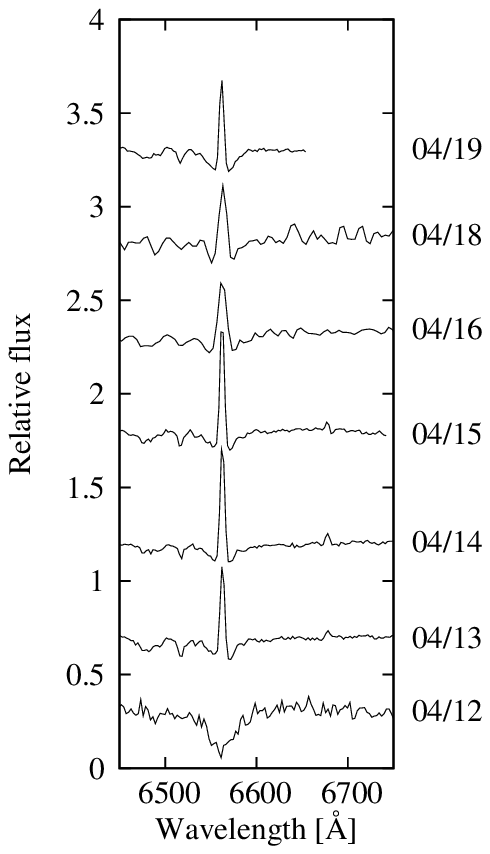}
  \end{center}
  \caption{
    
    Blow up of figure~\ref{fig:profile-variation}, showing the profiles
    from 4000 \AA~to 5000 \AA~(left) and around H$\alpha$ (right). Bottom
    corresponds to Period I and II, middle to Period III, top to Period
    IV.
    
  }
  \label{fig:Hd-HeI6678}
\end{figure*}

In this subsection, we explore the evolution of the structure of the
accretion disk in the superoutburst, mainly using the H$\alpha$
profile variation as described in section~\ref{section:RESULT}.  The
changes of the spectral features, shown in
figure~\ref{fig:Hd-HeI6678}, are summarized in
section~\ref{Summary of Spectral Change}.

\subsubsection{Summary of Spectral Change}
\label{Summary of Spectral Change}

We detected only Balmer absorption lines of H$\alpha$, H$\beta$, and H$\gamma$
(see figure~\ref{fig:Period-I}) in the rising phase of the
superoutburst. Around and just after the superoutburst maximum, 
the emission lines
of H$\alpha$, He~I~5876, He~I~6678, He~II~4686, and C~III/N~III were observed as well
as the absorption lines of Balmer components and He~I (see
figure~\ref{fig:Period-II}).
These high-excitation emission lines of He~II~4686 and C~III/N~III 
became much fainter by the fourth day from the superoutburst maximum.
In the latter part of the plateau, H$\alpha$ turned to an absorption 
line with a weak emission component, and He~I~6678 also changed into 
a weak absorption line (see figure~\ref{fig:Period-III}).
Other Balmer and He~I components remained to be absorption lines, 
and the Balmer components became deeper.
In the long fading tail, only emission lines of H$\alpha$, H$\beta$,
H$\gamma$, and He~I~6678 were observed (see figure~\ref{fig:Period-IV}).
The profile variability of He~I~6678 was almost the same as that of H$\alpha$
during our observations, except for April 12.

\subsubsection{Line Forming Mechanisms}

The absorption/emission profile of H$\alpha$ contains key information
to understand the evolution of the accretion disk
structure. Generally, an {\em absorption} line is formed when the
continuum emitter is optically thick with a normal temperature
gradient along the vertical axis (i.e., the temperature decreases
toward its surface).  On the other hand, an {\em emission} line can be
produced from (1) an optically thin, collisionally excited plasma, or
(2) a photoionized one, or (3) a temperature inversion layer of an
optically thick matter. In DNe, photoionization of an outer disk is
expected by irradiation from the white dwarf and the inner (hence
hotter) part of the accretion disk. The temperature inversion layer
may be formed on the surface of the disk by local heating by magnetic
activities or other mechanisms; in the following discussion, we do not
discuss the last possibility, however, since there is no observational
evidence for it in DNe at present.

\subsubsection{Similarity between GW Lib and WZ Sge}

For reference, we compare the spectral evolution of GW Lib observed in
our data, and that of WZ Sge during the 2001 superoutburst reported by
\citet{Baba2002} and \citet{Nogami2004}, the same type of DNe as GW
Lib seen with a high inclination. In both stars, the spectra showed
Balmer absorption lines in the rising phase.  Around and just after
the superoutburst maximum, H$\alpha$, He~II~4686, and C~III/N~III
turned to be emission lines in both systems. An asymmetric spiral
structure was seen in the doppler map of H$\alpha$ and He~II in WZ
Sge, which are considered to be connected with the early superhumps
\citep{Osaki2002, Kato2002}.  The region forming these emission lines
is located around the edge of the accretion disk in the Doppler map.

\subsubsection{Evolution of the Accretion Disk Structure}

Based on the comparison of spectral features between GW Lib and WZ Sge
during their superoutbursts, we propose the following scenario for the
evolution of the accretion disk that is commonly applicable to both
systems. From the results of WZ Sge, we consider that photoionization
is the most likely origin of the observed emission lines. If they were
instead produced from an optically thin, collisionally excited plasma,
the temperatures of those regions producing H$\alpha$ and He~II lines
would be different by factor $\sim$ 4. In reality,
however, the Doppler maps of WZ Sge show that both lines are emitted
at similar radii in the accretion disk
(\cite{Baba2002};\cite{Nogami2004}), where the temperature is expected
to be different by at most factor 2.
We note that this photoionized region is very
likely to be optically thin, considering the presence of the spiral
structure.

$\bullet${\it Rising phase}: The whole disk is optically thick and is
expanding in the rising phase of the superoutburst. This situation
means that irradiation from the white dwarf and the inner part of the
accretion disk was not yet effective at our first observation during
the rising phase. This is also consistent with the ``UV delay''
phenomenon that an increase of the ultraviolet flux lags that of the
optical flux by about 1 day (e.g. VW Hyi, \cite{Hassall1983}; SS Cyg,
\cite{Polidan1984}; WX Hyi, \cite{Hassall1985}; RX And,
\cite{Pringle1984}). The UV delay is interpreted to represent the time
the mass accreted from the outer edge of the accretion disk takes to
reach the white dwarf (\cite{Mineshige1988,Duschl1989,Meyer1989,
Livio1992,King1997}).

$\bullet${\it Maximum to plateau}: The disk then forms an outer extended
region, which starts to be strongly photoionized by the increasing UV
photons from the white dwarf and inner disk after around the
maximum. The photoionized region must be sufficiently large to subtend
a large solid angle, since the intensity of the emission component was
dominant in the H$\alpha$ profile around the maximum. While the
superoutburst is continuing in the plateau phase, the region shrinks
as the matter is accreted into smaller radii. Such geometrical
evolution of the accretion disk during the outburst is theoretically
expected (\cite{Osaki1996}) and is actually indicated by observations
of other DNe in eclipse (e.g. U Gem: \cite{Smak1984}; Z Cha:
\cite{O'Donoghue1986}; IP Peg: \cite{Wood1989}, and
\cite{Wolf1993}). The decrease of the UV photon flux also reduces the
intensity of the emission lines. The higher energy photons decreases
more rapidly, causing the faster decay of He~II emission lines
compared with H$\alpha$.

$\bullet${\it Fading phase}: In the long fading tail, the whole accretion disk
turns to be optically thin (without irradiation), and hence produces
emission lines by collisional excitation. This situation is the same
as in quiescence (\cite{Szkody2000}), although the accretion disk is
somewhat brighter, and hence the broad Balmer absorption features
produced by the white dwarf are apparently weaker, than in quiescence.

\subsubsection{Difference between GW Lib and WZ Sge }

A difference between GW Lib and WZ Sge is found in the H$\alpha$
component in the latter part of the plateau phase. In GW Lib,
all the emission lines of H$\alpha$, He~II~4686, and C~III/N~III
significantly weakened from the earlier epoch. In WZ Sge, while the
He~II~4686 and C~III/N~III lines weakened, the H$\alpha$ intensity
remained much stronger compared with GW Lib. The peak separations of
the He~II~4686 and C~III/N~III of WZ Sge became wider, indicating that
the region producing high excitation lines shrank into small
radii. This is consistent with our picture of the disk evolution. The
difference in the H$\alpha$ profile between the two stars may be
related to the rebrightening in WZ Sge after the main superoutburst,
which was not observed in GW Lib. The mechanism of the rebrightening
is still an open question.  Some authors, nevertheless, have suggested
that significant mass is left around the bright accretion disk
(e.g. \cite{Kato1998, Hellier2001b, Osaki2001}), and in fact
\citet{Uemura2008} found a large cool region around a hot accretion
disk by near-infrared observations of SDSS J102146.44+234926.3, a WZ
Sge-type DN, during the rebrightening phase. By contrast, GW Lib did
not have this region during the plateau phase, which could not cause a
rebrightening and a strong H$\alpha$ emission line.

\section{SUMMARY}
\label{section:SUMMARY}

\begin{enumerate}
  
\item We have compiled a large number of optical spectroscopic data of
  GW Lib taken by the international observation campaign during the 2007
  superoutburst. The observations started from the very beginning of the
  superoutburst, covering the rising phase, maximum, plateau, and long
  fading phase. We find that the profiles of spectral lines dramatically
  changed during the superoutburst.
  
\item Comparing our results with those of WZ Sge, the same type of DNe
  viewed with a high inclination angle, we construct a unified picture
  for the evolution of the accretion-disk structure in superoutburst
  applicable to these systems.
  
\item In the rising phase of the superoutburst, all Balmer lines are
  in absorption. We consider that the whole accretion disk is optically
  thick and is extending toward the white dwarf. The UV delay is
  consistently explained.
  
\item Around the maximum, the spectrum showed singly-peaked emission
  lines of H$\alpha$, He~I~6678, He~II~4686, and C~III/N~III superposed
  on absorption lines of Balmer components and He~I. We interpret that
  the emission lines are produced by photoionization of an optically
  thin region formed at the outer part of the disk, irradiated from the
  white dwarf and the inner part of the disk. The emission components
  significantly weakened in the latter part of the plateau phase. This
  can be explained by the shrinkage of the optically-thin region due to
  accretion and by decrease of irradiation.
  
\item In the fading tail, all Balmer lines were emission lines, as
  observed in quiescence. We consider that the whole disk is optically
  thin and emits lines by collisional excitation in this state.
  
\end{enumerate}

\bigskip

The authors are thankful to amateur observers for continuously
reporting their valuable observations to VSNET.  We are also indebted
to Taichi Kato for his useful comments on the draft.  This work is
partly supported by Research Fellowships of Japan Society for the
Promotion of Science for Young Scientists (AI), and by the
Grant-in-Aid for the Global COE Program ``The Next Generation of
Physics, Spun from Universality and Emergence'' from the Ministry of
Education, Culture, Sports, Science and Technology (MEXT) of Japan.

\end{document}